# Big Plastic Masses Detection using Sentinel 2 Images

Fernando Martín-Rodríguez.

fmartin@tsc.uvigo.es.
atlanTTic research center for Telecommunication Technologies, University of Vigo,
Campus Lagoas Marcosende S/N, 36310 Vigo, Spain.

*Abstract-* **This communication describes a preliminary research on detection of big masses of plastic (marine litter) on the oceans and seas using EO (Earth Observation) satellite systems. Free images from the Sentinel 2 (Copernicus Project) platform are used. To develop a plastic recognizer, we start with an image where we can find a big accumulation of "non-floating" plastic: Almería greenhouses. We made a test using remote sensing differential indexes, but we got much better results using all available wavelengths (thirteen frequency bands) and applying Neural Networks to that feature vector.**

## I. Introduction

Marine litter has become a major environmental problem today. Consequently, there are many initiatives to fight this problem from different areas. Part of these initiatives, which we could call "proximity projects", focus efforts on beaches and/or in shallow waters near the coastline [1,2].

Other initiatives aim to attack the problem of the large masses of floating garbage that accumulate in the sea. It is known that currents and tides tend to accumulate waste in large masses that can be considered true islands of garbage. Although marine litter can be classified into many different categories [3], the most populated class (and the most important in the case of floating debris) is plastic litter.

In the latter case of large floating accumulations, the use of satellite image capture technologies (EO: Earth Observation) is interesting. Currently, interest is in developing systems that take advantage of available satellites, especially those of the Copernicus project [4,5].

In particular, the Sentinel 2 satellite produces multi-spectral images with thirteen wavelengths (or frequency bands) that can surely be useful. Sentinel 2 public images have a precision of 10m/px (on visible bands), which is certainly low but sufficient for large masses.

The development of applications of this type can help efforts against marine litter by detecting large accumulations, quantifying and monitoring them. When there are suspicions of accumulation points due to currents or tides, it can be used to confirm or deny them.

Sentinel 2 is limited to near-shore waters (although it includes the entire Mediterranean Sea). In the ocean, we could extend development to the most recent Sentinel 3 (21 bands, 300m/px).

For this development, it is necessary to have images with significant amounts of plastic material. Although it is not floating plastic, an area known for huge plastic covers is located in Almería, Spain (where there exist enormous extensions of plastic greenhouses, figure 1).

TABLE I
Available bands in Sentinel 2, MSI sensor [6]

| Bands | Wavelength (nm) | Resolution (m) |
|---|---|---|
| B1 (Coastal aerosol) | 443 | 60 |
| B2 - Blue | 490 | 10 |
| B3 - Green | 560 | 10 |
| B4 - Red | 665 | 10 |
| B5 VNIR/RedEdge | 705 | 20 |
| B6 VNIR/RedEdge | 740 | 20 |
| B7 VNIR/RedEdge | 783 | 20 |
| B8 - NIR | 842 | 10 |
| B8A VNIR/RedEdge | 865 | 20 |
| B9 Water vapour | 945 | 60 |
| B10 SWIR Cirrus | 1375 | 60 |
| B11 - SWIR | 1610 | 20 |
| B12 - SWIR | 2190 | 20 |

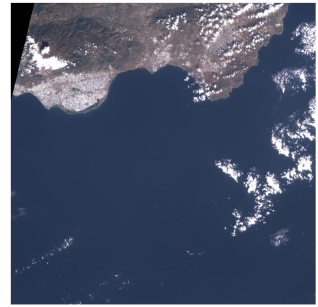

Fig. 1. RGB composition from Sentinel 2, Almería, 100 Mpx, 10 m/px (100x100 Km$^2$).

## II. Normalized Indexes

In remote sensing, the so-called normalized indexes are often used. These indexes are computed from pairs of chromatic components [7,8].

An index based methodology is presented in [13], where authors recommend a combination of two indexes: the well-known NDVI (vegetal masses detection) and novel FDI (Folating Debris Index). Testing this method on the former image, we get the result in figure 2.

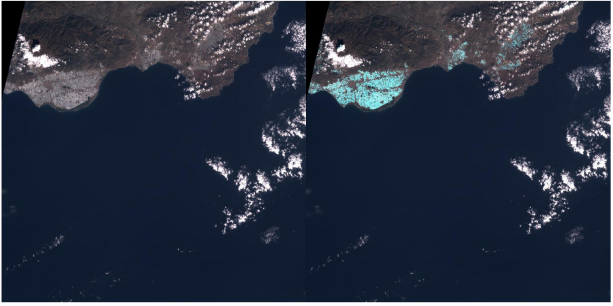

Fig. 2. Result using combination of indexes from [13].

Studying plastic behavior at different wavelengths, we thought about an alternative index using bands 9 and 8. Id EST: computing (B8-B9)/(B8+B9). Result is shown in figure 3, this new option shows more probability of false detection (less missed detections).

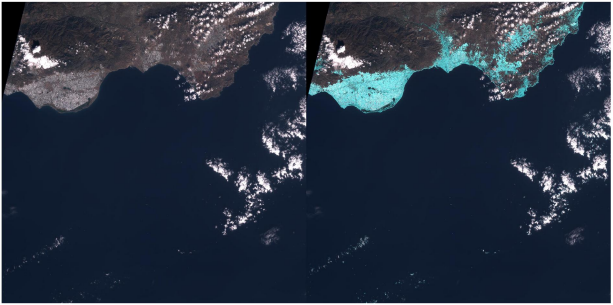

Fig. 3. Result of bands 8 & 9 index.

Combining these methods with a cloud detection method [14] and discarding the land areas with the help of maps, we could meet the purpose. However, in next section we will see more powerful tools that get better and more direct results.

### III. USE OF ARTIFICIAL NEURAL NETWORKS

With thirteen numerical values per pixel, we can think of this as a pattern recognition problem. For each point, we naturally obtain a thirteen components feature vector.

Here arises the idea of using artificial neural networks [9] to learn the implicit relationships between different bands that may characterize plastic.

#### A. Network Structure

In this case, we have a feature vector per pixel made up of thirteen (twelve-bit) integers (Sentinel format). There are only two classes (plastic and non-plastic), so a single output network will be enough to be trained to obtain values of 1.0 for plastic and 0.0 otherwise. Under these conditions a multilayer perceptron structure (MLP [9]) may work well.

We will have just one hidden level (three layers). We started testing with ten hidden neurons (slightly less than the number of inputs). Given the good results, we do not consider necessary going to more complex structures.

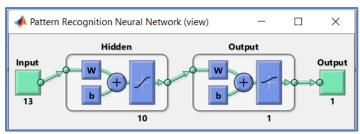

Fig. 4. Network structure.

Note that different bands also have different resolutions. To solve this difficulty, we have used two-dimensional interpolation type "Lanczos3" [10].

Since the network operates at pixel level, once trained, it can be used over any resolution.

#### B. Training, Validation and Test

For training, we have started by labeling the original image manually (drawing a detailed mask on top of the plastic areas, figure 5).

The number of samples is more than enough. Even reducing ourselves to the resolution of band 1, we have 3,348,900 samples for: 14 * 10 + 11 = 151 weights. Normally it is supposed that the number of samples should be over fifteen times the number of weights. That condition here is fulfilled widely.

We have balanced the training samples (dropping randomly some samples of the majority class to avoid training problems). Training was done using MATLAB [11] and back-propagation algorithm (conjugate gradient) [12]. 70% of the samples have been dedicated to training, 15% to validation (verification for algorithm completion) and 15% to final testing.

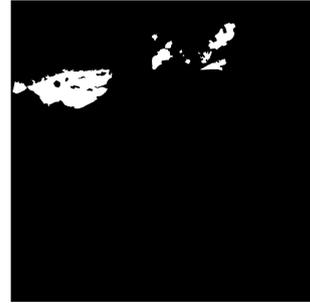

Fig. 5. Truth table.

Fig. 6. Confusion matrixes.

The training has been successful, obtaining an error rate of 2% on the test set. See figure 6 (confusion matrixes) and figure 7: recognition result on a **different** image of the same place (to demonstrate generality of method).

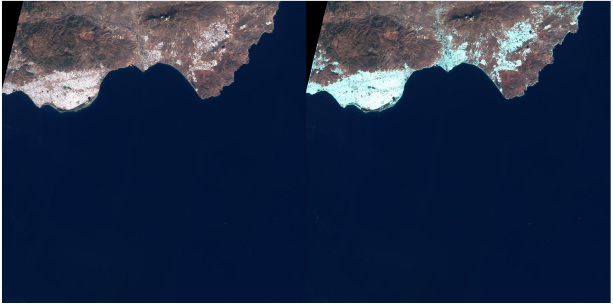

Fig. 7. Neural network result.

## IV. CONCLUSIONS AND FUTURE LINES

We have studied methods for the automatic detection and quantification of plastic waste floating in the sea. Although the use of differential indexes (very typical in remote sensing) may have some utility, the use of neural networks on all the frequency bands is the solution that provides the best result.

This system can be used to quickly confirm or discard places suspected of accumulating waste. In the future, it is planned to develop an application that automatically downloads and analyzes images of all available marine areas (using cartographic information to discard land), detecting, quantifying and monitoring large accumulations.

Note that Sentinel 2 images have an average periodicity of two days and a half.

To study the great oceans, the work would have to be extended to the OLCI sensor of Sentinel 3 (21 bands). In principle, a retraining of the neural networks should be enough (another option would be to choose the 13 bands that are closest in wavelength to those contemplated in this study).

## V. ACKNOWLEDGEMENTS

We are grateful for the collaboration of the "Fundación Biodiversidad" (Biodiversity Foundation) of the "Ministerio para la Transición Ecológica y el Reto Demográfico" (Ministry for the Ecological Transition and Demographic Challenge) through the "Pleamar-2019" program, co-financed by the EMFF (European Maritime and Fisheries Fund).